\documentstyle[12pt,aaspp4,flushrt]{article}

\def\ffrac#1#2{{\textstyle\frac{#1}{#2}}}

\def\p{\partial}
\def\bfp{{\bf p}}
\def\bfq{{\bf q}}
\def\bfr{{\bf r}}
\def\bfv{{\bf v}}

\def\bfz{{\bf z}}

\def\bfM{{\bf M}}
\def\bfB{{\bf B}}
\def\bfX{{\bf X}}
\def\bfY{{\bf Y}}
\def\bfS{{\bf S}}
\def\bfD{{\bf D}}
\def\bfK{{\bf K}}
\def\bfT{{\bf T}}
\def\bfA{{\bf A}}

\def\bfP{{\bf P}}
\def\bfQ{{\bf Q}}
\def\bfZ{{\bf Z}}
\def\bfC{{\bf C}}

\def\bfL{{\bf L}}
\def\bfN{{\bf N}}
\def\bnabla{\mbox{\boldmath $\nabla$}}
\def\sia{{\scshape sia}}
\def\sias{{\scshape sia}s}

\def\half{{\textstyle{1\over2}}}
\def\be{\begin{equation}}
\def\ee{\end{equation}}
\def\halff{{1/2}}
\def\mathnew{\mathsurround=0pt}
\def\simov#1#2{\lower .5pt\vbox{\baselineskip0pt \lineskip-.5pt
        \ialign{$\mathnew#1\hfil##\hfil$\crcr#2\crcr\sim\crcr}}}

\begin{document}

\title{A class of symplectic integrators with adaptive timestep for separable
Hamiltonian systems}

\author{Miguel Preto and Scott Tremaine}

\medskip

\affil{Princeton University Observatory, Peyton Hall, Princeton, 
NJ 08544, USA}

\medskip

\abstract{Symplectic integration algorithms are well-suited for long-term
integrations of Hamiltonian systems because they preserve the geometric
structure of the Hamiltonian flow. However, this desirable property is
generally lost when adaptive timestep control is added to a symplectic
integrator. We describe an adaptive-timestep symplectic integrator that can be
used if the Hamiltonian is the sum of kinetic and potential energy components
and the required timestep depends only on the potential energy
(e.g. test-particle integrations in fixed potentials). In particular, we
describe an explicit, reversible, symplectic, leapfrog integrator
for a test particle in a near-Keplerian potential; this integrator has
timestep proportional to distance from the attracting mass and has the
remarkable property of integrating orbits in an inverse-square force field
with only ``along-track" errors; i.e. the phase-space shape of a Keplerian
orbit is reproduced exactly, but the orbital period is in error by
O$(N^{-2})$, where $N$ is the number of steps per period.}

\section{Introduction}

During the last decade a great deal of effort has been devoted to the
development of symplectic integration algorithms (\sias) for
Hamiltonian systems (\cite{CS90,Yo93,MPS96}).  An \sia\ is a symplectic
mapping of phase space $\bfz=(\bfq,\bfp)$ and time, $\bfM_h:(\bfz,t)\to
(\bfz',t'=t+h)$, that approximates the Hamiltonian flow over a
small interval $h$.  {\scshape Sia}s preserve much of the
geometric structure of the Hamiltonian flow; as a result, they usually
have only oscillatory and not secular errors in the integrals of
motion and are useful when the main goal is minimizing long-term
qualitative errors rather than achieving the highest possible
short-term precision.

The most popular \sia\ is the leapfrog or Verlet method, which can be
applied to separable Hamiltonians of the form 
\be
H(\bfq,\bfp,t)=T(\bfp)+U(\bfq,t) 
\label{eq:hamsep}
\ee 
(usually $T$ and $U$ are the kinetic and potential energy). We may
define ``drift'' and ``kick'' operators
\begin{equation}
\bfD_h:(\bfq,\bfp,t)\to\left(\bfq+h\frac{\p T}{\p\bfp},\bfp,t+h\right), \qquad
\bfK_h:(\bfq,\bfp,t)\to\left(\bfq,\bfp-h\frac{\p U}{\p\bfq},t\right),
\label{eq:kd}
\end{equation}
and the leapfrog operator is then
\begin{equation}
\bfL_h=\bfD_{h/2}\bfK_h\bfD_{h/2}, \quad \hbox{or} \quad
\bfL_h=\bfK_{h/2}\bfD_h\bfK_{h/2};
\label{eq:leapdef}
\end{equation}
either of which defines a second-order \sia\ 
(``DKD leapfrog'' and
``KDK'' leapfrog). Higher-order \sias\ can be derived by concatenating
leapfrog operators; for example
$\bfL_{x_1h}\bfL_{x_0h}\bfL_{x_1h}$ is a fourth-order \sia\
if $x_0=-2^{1/3}/(2-2^{1/3})$, $x_1=1/(2-2^{1/3})$
(\cite{Yo90}). Leapfrog and its higher-order generalizations have
several appealing features: only a small number of force evaluations
is required per step (1 for a second-order integrator, 3 for a
fourth-order integrator); no auxiliary variables are required, thus
minimizing memory requirements; and the integrators are explicit and 
time-reversible.

One limitation of \sias\ is that they are usually restricted to a fixed
timestep. When a standard adaptive-timestep prescription is applied to an
\sia, its performance is no better than that of non-symplectic integrators;
the reason is that the mapping $\bfM_{h(\bfz)}(\bfz,t)$ is not
symplectic even when $\bfM_h(\bfz,t)$ is. This is a serious limitation, since
in most applications a fixed timestep is inefficient. 

There have been many attempts to construct adaptive-timestep \sias. In
the context of molecular dynamics, Skeel \& Biesiadecki (1994)
\nocite{SB94} split the interaction potential into a short-range
component (rapidly varying, cheap to calculate, zero outside a limited
range) and a long-range component (slowly varying, expensive to
calculate); they then evaluate the effects of the short-range
potential every timestep using a symplectic integrator, adding in $N$
times the long-range potential at every $N^{\rm th}$ timestep. This
procedure retains symplecticity and can be generalized by splitting
the potential into any number of parts. Duncan, Levison \& Lee (1998)
\nocite{DLL98} describe a variant of the Skeel-Biesiadecki method for
long-term integrations of solar-system orbits.

In some situations an adaptive-timestep integrator can be replaced by an
integrator that uses different---but constant---timesteps for
different subsystems. In the context of solar system integrations,
Saha \& Tremaine (1994) \nocite{ST94}\ describe an \sia\ that uses a
different timestep for each planet, which works well so long as the
planetary orbits are well-separated.

An alternative to adaptive-timestep \sias\ is to abandon symplecticity but
demand that the integrator is time-reversible. Formally, an integrator is
reversible if $\bfM_h\bfT\bfM_h=\bfT$ where $\bfT$ is the time-reversal
operator. Reversible maps have many of the same geometric properties as
symplectic maps (Arnold 1984 \nocite{Ar84} calls the similarities
``astonishing'') and hence we expect that reversible integration algorithms
have similar virtues to \sias\ when they are applied to reversible Hamiltonian
systems. In particular we expect that reversible methods should not exhibit
secular errors in the integrals of motion.

Reversible integration algorithms with adaptive timestep are relatively easy
to generate.  Any integration algorithm $\bfN_h$ can be converted to many
reversible algorithms of the same order (\cite{HFK97}); one example is
$\bfM_h\equiv \bfN_{h/2}\bfT\bfN^{-1}_{h/2}\bfT$.  Moreover any reversible
integration algorithm remains reversible with variable timestep if the
timestep depends symmetrically on the initial and final phase-space
coordinates (\cite{HMM95}); unfortunately such integrators are usually
implicit and therefore slower than explicit integrators. Various explicit
reversible integrators with adaptive timestep are described by Huang \&
Leimkuhler (1997), Quinn et al. (1997), Calvo et al. (1998), and Evans \&
Tremaine (1999). \nocite{HL97,CLMSS98,Qu97,ET99}

One problem that requires adaptive timestep is the integration of highly
eccentric near-Keplerian orbits (e.g.  long-period comets, which often have
eccentricities $e>0.99995$). Here the standard approach is to regularize the
equations of motion; in particular the Kustaanheimo-Stiefel (KS)
regularization converts the Keplerian Hamiltonian to a harmonic oscillator
Hamiltonian, which is easier to integrate numerically.  KS regularization can
be extended to handle few-body problems but is restricted to inverse-square
interparticle forces.  This technique is also useful in simulations of star
clusters, to deal with the delicate problem of the formation and dynamical
evolution of tightly-bound binary stars over very long times
(e.g. \cite{MA93}). Mikkola (1997, see also \cite{RH99}) \nocite{Mi97} has
combined KS regularization with an efficient \sia\ designed for nearly
integrable problems by Wisdom \& Holman (1991) \nocite{WH91} to provide a
sophisticated integrator for eccentric near-Keplerian orbits.

A closely related problem in numerical celestial mechanics is the long-term
integration of moderate-eccentricity planet-crossing orbits
(e.g. Earth-crossing asteroids, Centaurs, Jupiter-family comets), which are
nearly Keplerian for millions of years, yet occasionally suffer strong
perturbations from close planetary encounters that may only last a few 
hours (\cite{DLL98}).

The aim of this paper is to discuss a class of explicit adaptive-timestep
\sias\ that can be used to follow Hamiltonian systems of the form
(\ref{eq:hamsep}) in the important special case where the timestep depends
only on the potential energy $U(\bfq,t)$. We provide a brief review of \sias\
in \S \ref{sec:review}, and discuss the use of extended phase space to derive
adaptive-timestep \sias\ in \S \ref{sec:extend}. A class of explicit
adaptive-timestep \sias\ that are particularly suitable for following orbits
in nearly Keplerian potentials is presented in \S\ref{sec:keptwo}, and
numerical tests are described in \S \ref{sec:num}. Section \ref{sec:disc}
contains a brief summary.

As this work neared completion, we learned of a paper by Mikkola \& Tanikawa
(1999) \nocite{MT99} that contains many similar conclusions.

\section{Review of symplectic integration}

\label{sec:review}

A \sia\ is a mapping of the form
$\bfM_h:(\bfz,t)\to(\bfz',t'=t+h)$
where $\bfM_h$ is symplectic and $\bfz'$ approximates the phase-space 
trajectory of $\bfz$ over time $h$, which is given by
\begin{equation}
\dot\bfz=\{\bfz,H\}, 
\label{eq:hamilton}
\ee where $H(\bfz,t)$ is the Hamiltonian and the
braces stand for the Poisson bracket.

The \sia\ can be defined implicitly by a generating function
$W=W(\bfq,\bfp',t)$, with the equations of transformation
\begin{equation}
\bfp = \frac{\p{W}}{\p{\bfq}}, \qquad \bfq' =
\frac{\p{W}}{\p{\bfp'}}. 
\end{equation}
In the simplest \sia, the generating function is chosen to be 
\begin{equation}
W=\bfq \cdot \bfp' + h H(\bfq,\bfp',t),
\label{eq:simple}
\end{equation}
which implies
\begin{eqnarray}
\bfq' & = & \bfq + h \frac{\p{H(\bfq,\bfp',t)}} {\p{\bfp'}}, \nonumber \\
\bfp' & = & \bfp - h \frac{\p{H(\bfq,\bfp',t)}} {\p{\bfq}}. 
\end{eqnarray}
These equations define an implicit first-order \sia. Higher order schemes can
be derived from more complicated generating functions (e.g. \cite{CS90}).

We can go further if the Hamiltonian is separable and autonomous, that is, if 
\begin{equation}
H=H_{A}+H_{B}
\end{equation}
where $H_{A}$ and $H_{B}$ are time-independent and separately integrable. For a
system of this type, the equations of motion can be written as
\begin{equation}
\dot\bfz = \{ \bfz, H_{A}+H_{B} \}.
\label{eq:hamm}
\end{equation} 
We introduce the differential operators $\bfA=\{\cdot
,H_{A}\}$, $\bfB=\{\cdot ,H_{B}\}$, and write the formal solution of equation
(\ref{eq:hamm}) as 
\begin{equation}
\bfz(t) = \exp [t(\bfA+\bfB)] \bfz(0).
\end{equation} 
By assumption, we know how to calculate $\exp(t\bfA)$ and $\exp(t\bfB)$. In
general, these operators are non-commutative so
$\exp[t(\bfA+\bfB)]\not=\exp(t\bfA)\exp(t\bfB)$.  The correct expression is
given by the Baker-Campbell-Hausdorff (BCH) identity (\cite{Yo93}),
\begin{equation}
\exp(\bfX)\exp(\bfY) = \exp(\bfZ),
\end{equation}
where $\bfZ=\bfX+\bfY+\frac{1}{2}[\bfX,\bfY]+
\frac{1}{12}[\bfX-\bfY,[\bfX,\bfY]]+\cdots$.  To construct an explicit
symplectic integrator of order $n$ we use the BCH identity to find a set of
real numbers $(c_{i},d_{i})$ such that
\begin{equation}
\exp[h(\bfA+\bfB)] = \prod_{i=1}^{k} \exp(c_{i}h\bfA) \exp (d_{i}h\bfB) +
O(h^{n+1});
\end{equation}
the integrator is then
\begin{equation}
\bfz(t=h)= \left[ \prod_{i=1}^{k} \exp(c_{i}h\bfA) \exp
(d_{i}h\bfB)\right]\bfz(0), 
\end{equation} 
where the operators are applied in the order
$(c_1,d_1,\ldots,c_k,d_k)$; this is an example of the general
technique of operator splitting.  See Yoshida (1990) \nocite{Yo90} for
a systematic strategy of finding these numerical coefficients. In the
important case of a Hamiltonian of the form (\ref{eq:hamsep}), this
map takes the simple form (cf. eq. \ref{eq:kd})
\be
\bfq_{i+1} =  \bfq_{i} + h c_{i}
       \left(\frac{\p{T}}{\p{\bfp}}\right)_{\bfp_{i}}, \qquad
\bfp_{i+1} = \bfp_{i} - h d_{i}
       \left(\frac{\p{U}}{\p{\bfq}}\right)_{\bfq_{i+1}},
\ee
where $\bfz(0)=\bfz_1$, $\bfz(h)=\bfz_{k+1}$. 
The usual second-order leapfrog (eq. \ref{eq:leapdef}) 
corresponds to the choice $k=2$, $c_{1}=c_{2}=\frac{1}{2}$, $d_{1}=1$ and
$d_{2}=0$ for DKD leapfrog (or $c_1=0$,
$c_2=1$, $d_1=d_2=\half$ for KDK leapfrog).
Using the BCH identity, one can show that
\begin{equation}
\exp(\half h\bfA) \exp(h\bfB) \exp(\half h \bfA)=\exp[h(\bfA+\bfB+\bfC)]
\quad \hbox{where} \quad \bfC=\{\cdot, H_{\rm err}\}
\end{equation}  
and 
\begin{equation}
H_{\rm err} = \frac{h^{2}}{12}\{\{H_{A},H_{B}\}, H_{B}+\half H_{A}\} +
O(h^{4}).
\label{eq:uuuuu}
\end{equation}
Thus DKD leapfrog describes the equations of motion in a
surrogate Hamiltonian 
\begin{equation}
\widetilde{H} = H + H_{\rm err};
\end{equation}
(more precisely the numerical trajectory lies exponentially close to the exact
trajectory of the surrogate Hamiltonian). The good properties of symplectic
integrators, such as the absence of secular errors in the energy, are a
consequence of the existence of this Hamiltonian.

\section{Variable timestep and extended phase space}

\label{sec:extend}

We want to construct an explicit adaptive-timestep \sia\ to integrate the
Hamiltonian (\ref{eq:hamsep}).
Following Mikkola (1997), \nocite{Mi97} we extend phase space
by introducing a fictitious time variable $\tau$ through the relation
\begin{equation} 
dt = g(\bfq,\bfp,t) d\tau 
\end{equation} and take $t\equiv
q_0$ as a new coordinate together with the corresponding conjugate momentum
$p_{0}=-H$. Thus an extended phase space is defined by 
\begin{equation}
\bfQ=(q_{0},\bfq), \qquad \bfP=(p_{0},\bfp), 
\end{equation} and the equations
of motion in the extended phase space are 
\begin{eqnarray} 
\frac{d\bfQ}{d\tau}
& = & g(\bfq,\bfp,t)\frac{\p{H}}{\p{\bfP}}=\frac{\p{\Gamma}}{\p{\bfP}},
\nonumber \\ \frac{d\bfP}{d\tau} &= & -g(\bfq,\bfp,t)\frac{\p{H}}
{\p{\bfQ}}=-\frac{\p{\Gamma}}{\p{\bfQ}}, 
\label{eq:eqmot}
\end{eqnarray} 
where 
\begin{equation}
\Gamma(\bfQ,\bfP)=g(\bfq,\bfp,q_0)[H(\bfq,\bfp,q_0)+p_{0}], 
\label{eq:gamma}
\end{equation} 
and only trajectories on the hypersurface $\Gamma=0$ in the extended phase
space correspond to solutions of the equations of motion in the original phase
space. The equations of motion (\ref{eq:eqmot}) are Hamiltonian in the
extended phase space if $\Gamma$ is chosen as the Hamiltonian. We can now
integrate the equations of motion with an \sia\ having constant fictitious
timestep $\Delta{\tau}$ in the extended phase space, which is equivalent to a
variable timestep $\Delta{t}=g(\bfq,\bfp,t)\Delta\tau$ in the reduced phase
space.

The Hamiltonian (\ref{eq:gamma}) is not generally separable and hence
operator-splitting techniques cannot be used to derive explicit \sias\ with
arbitrary timesteps.  Nevertheless, this approach can yield useful \sias\ for
specific choices of the timestep function $g(\bfq,\bfp,t)$.

\subsection{A separable Hamiltonian in extended phase space}

\label{sec:sep}

We choose the timestep function to be 
\begin{equation}
g(\bfQ,\bfP)=\frac{f(T_e(\bfP))-f(-U(\bfQ))}{T_e(\bfP)+U(\bfQ)},
\label{eq:gdef}
\end{equation}
where  $T_e(\bfP)=T(\bfp)+p_{0}$ and $U(\bfQ)=U(\bfq,q_0)$. The
Hamiltonian (\ref{eq:gamma}) becomes 
\begin{equation}
\Gamma(\bfQ,\bfP)=f(T_e(\bfP))-f(-U(\bfQ)),
\label{eq:xxx}
\end{equation}
which is separable. The equations of motion are
\begin{eqnarray}
         {d{q}_{i}\over d\tau}=\frac{\p{\Gamma}}{\p{p_{i}}} & = & 
f^{\prime}(T(\bfp)+p_0)\frac{\p T(\bfp)}{\p p_{i}}, \nonumber \\
{d t\over d\tau}= \frac{\p{\Gamma}}{\p{p_{0}}} & = & 
f^{\prime}(T(\bfp)+p_0), \nonumber \\
         {d {p}_{i}\over d\tau}=-\frac{\p{\Gamma}}{\p{q_{i}}} & = & 
-f^{\prime}(-U(\bfq,t))\frac{\p U(\bfq,t)}{\p q_{i}}, \nonumber \\
         {d{p}_{0}\over d\tau}=-\frac{\p{\Gamma}}{\p{q_{i}}} & = & 
-f^{\prime}(-U(\bfq,t))\frac{\p U(\bfq,t)}{\p t}.
\label{eq:metaeqns}
\end{eqnarray}

To choose the function $f$ we recall that
$H(\bfq,\bfp,q_0)+p_{0}=T_e(\bfP)+U(\bfQ)=0$ for the Hamiltonian
flow, i.e., $T_e(\bfP)\simeq{-U(\bfQ)}$ during the numerical
integration. Consequently, $f(T_e)-f(-U) \simeq 0$ and we can Taylor expand the
function $f$ around $T_e(\bfP)$ to obtain
\begin{equation}
 f(T_e(\bfP))=f(-U(\bfQ))+[T_e(\bfP)+U(\bfQ)]f'(-U(\bfQ))+\hbox{O}(T_e+U)^2.  
\end{equation}
Therefore, equation (\ref{eq:gdef}) yields 
\begin{equation}
g(\bfQ,\bfP)\simeq f'(-U)
\label{eq:fff}
\end{equation}
along the integration path. Thus the timestep can be chosen to be an
arbitrary function of the potential energy, $g=g(-U)$, and a suitable $f(-U)$
is determined by integrating $g(-U)$.  

The choice of the timestep function is crucial to the success of an
integrator, and the restriction that this function depends only on the
potential energy $U$ is severe. Nevertheless timestep functions of this
form can be useful for a variety of dynamical problems. 

\subsection{Error analysis}

We now illustrate how to analyze the integration errors that arise when
fixed-timestep \sias\ are used to integrate the equations of motion
(\ref{eq:metaeqns}) in extended phase space.  For simplicity we shall assume
that the potential is stationary, $U(\bfQ)=U(\bfq)$, and restrict our attention
to DKD leapfrog.

The error Hamiltonian for DKD leapfrog applied to the Hamiltonian
(\ref{eq:xxx}) is 
\begin{eqnarray} 
\lefteqn{\Gamma_{\rm err}(\bfQ,\bfP) = 
\ffrac{1}{12}(\Delta\tau)^2\left[f'(-U)\right]^2U_{,i}U_{,j}
\left[f''(T_e)p_ip_j+f'(T_e)\delta_{ij}\right]}\nonumber \\
& {} & -\ffrac{1}{24}(\Delta\tau)^2
\left[f'(T_e)\right]^2p_ip_j\left[-f''(-U)U_{,i}U_{,j} 
+ f'(-U)U_{,ij}\right] + \hbox{O}(\Delta\tau)^4,
\label{eq:gamerr}
\end{eqnarray}
where $U_{,i}=\p U/\p q_i$ and summation over the indices $i,j\in\{1,2,3\}$ is 
assumed.

Once we have the error Hamiltonian, the numerical error in the energy in the
original phase space is easy to derive. The integrator accurately follows the
trajectory of the surrogate Hamiltonian
\be
\widetilde\Gamma= \Gamma +\Gamma_{\rm err}=f(T_e)-f(-U) +\Gamma_{\rm err};
\label{eq:gamm}
\ee
thus $\widetilde\Gamma$ is conserved along the numerical trajectory. At the
starting point $(\bfQ_i,\bfP_i)$, $\Gamma=0$ so
$\widetilde\Gamma=\Gamma_{\rm err}(\bfQ_i,\bfP_i)\equiv \Gamma_i$ 
throughout the integration.
Since $\widetilde\Gamma$ is independent of the coordinate $q_0$, 
the momentum $p_0$ is conserved
throughout the integration and is therefore equal to minus the initial energy 
$E_i$. Thus $T_e=T(\bfp)+p_0=\Delta E-U(\bfq)$ where $\Delta E=E-E_i$ is the
energy error. Equation (\ref{eq:gamm}) can now be rewritten as
\be
\Gamma_i=f(\Delta E-U(\bfq))-f(-U(\bfq))+
\Gamma_{\rm err}(\bfQ,\bfP);
\ee
since the energy error is small we can expand in a Taylor series to obtain
\be
\Delta E={\Gamma_i-\Gamma_{\rm err}(\bfQ,\bfP)\over
f'(-U(\bfq))}.
\label{eq:enerr}
\ee

\section{Keplerian two-body problem}

\label{sec:keptwo}

The long-term integration of nearly Keplerian orbits is central to the study
of solar system dynamics, and the Keplerian two-body problem provides a
natural laboratory for testing integration algorithms.

For simplicity we work in two dimensions, and to provide more general formulae
we add an extra potential $V(\bfq,t)$ to the point-mass potential that defines
the Keplerian problem. The Hamiltonian for a test particle is thus
\be 
H(\bfq,\bfp,t)=\half \bfp^2-{\mu\over r}+V(\bfq,t), 
\ee 
where $\bfq=(x,y)$, $\bfp=\bfv=(v_x,v_y)$, $r^2=x^2+y^2$, and $\mu$ is 
the mass. The equations of motion are 
\begin{equation}
\frac{d^2x}{dt^2}={dv_x\over dt}=-\mu\frac{x}{r^{3}}-\frac{\p V}{\p x},\qquad
\frac{d^2y}{dt^2}={dv_y\over dt}=-\mu\frac{y}{r^{3}}-\frac{\p V}{\p y}. 
\end{equation}

There are two natural choices for the timestep function $g(\bfq,\bfp)$
when integrating bound Keplerian orbits: (i) $g\propto r^{3/2}$
ensures that the timestep is a constant fraction of the local
free-fall time $t_{\rm ff}\sim r^{3/2}\mu^{-1/2}$, so that all phases of
highly eccentric orbits are followed with the same relative accuracy;
(ii) $g\propto r$ ensures that the coordinate trajectory as a function
of the fictitious time is that of a harmonic oscillator, so
there are no high-frequency harmonics that are difficult for numerical
integrators to follow. 

We can accommodate both of these choices by assuming that the timestep
function is a power law in radius, 
\be 
g(r)=\epsilon r^{\gamma}\mu^{1-\gamma}, 
\label{eq:gpow}
\ee 
where $\epsilon$ is a constant that parametrizes the size of the timestep;
having introduced $\epsilon$ we can henceforth set $\Delta\tau=1$ without loss
of generality. Since $U\simeq-\mu/r$ for nearly Keplerian orbits, equation
(\ref{eq:fff}) then suggests that we choose
\be
f(x) = \left\{ \begin{array}{ll} \frac{\displaystyle 
\epsilon\mu}{\strut \displaystyle 1-\gamma} x^{-\gamma+1} 
& \mbox{if $\gamma \neq 1,$} \\
\epsilon\mu \log x & \mbox{if $\gamma = 1$.} 
\end{array}
\right.
\label{eq:fdef}
\ee
The corresponding Hamiltonian is
\be
\Gamma(\bfQ,\bfP) = \left\{ \begin{array}{ll} \frac{\displaystyle
\epsilon\mu}{\strut \displaystyle 1-\gamma} 
\left([T_e(\bfP)]^{-\gamma+1}-[-U(\bfQ)]^{-\gamma+1}\right) & \mbox{if
$\gamma\neq 1$} \\ \epsilon\mu \left(\log [T_e(\bfP)]-\log [-U(\bfQ)] \right) 
& \mbox{if $\gamma = 1$.} \end{array}
\right.
\ee
For the problem we consider here, $U(\bfQ)=-\mu/(x^2+y^2)^{1/2}+
V(x,y,t)$ and $T_e(\bfP)=\frac{1}{2}(v_x^2+v_y^2)+p_{0}$. The equations of 
motion (\ref{eq:metaeqns}) in the fictitious time read
\begin{eqnarray}
        {d x\over d\tau} & = &\epsilon\mu\frac{v_{x}}{(\frac{1}{2}{v_{x}}^{2}+
\frac{1}{2}{v_{y}}^{2}+p_{0})^{\gamma}} \nonumber         \\
        {d y\over d\tau} & = &\epsilon\mu\frac{v_{y}}{(\frac{1}{2}{v_{x}}^{2}+
\frac{1}{2}{v_{y}}^{2}+p_{0})^{\gamma}} \nonumber         \\
        {d t\over d\tau} & = & \epsilon\mu\frac{1}{(\frac{1}{2}{v_{x}}^{2}+
\frac{1}{2}{v_{y}}^{2}+p_{0})^{\gamma}}, \nonumber \\
   {d v_x\over d\tau}& = &-{\epsilon\mu\over (\mu/r-V)^\gamma}\left({\mu x\over
r^3}+{\p V\over \p x}\right), \nonumber \\
   {d v_y\over d\tau}& = &-{\epsilon\mu\over (\mu/r-V)^\gamma}\left({\mu y\over
r^3}+{\p V\over \p y}\right), \nonumber \\
   {d p_0\over d\tau}& = &-{\epsilon\mu\over (\mu/r-V)^\gamma}{\p V\over \p t},
\label{eq:ffff}
\end{eqnarray}
These equations can be integrated using leapfrog, since
the right-hand sides of the first three depend only on momenta in the extended 
phase space, and the latter three on coordinates. For example, if the extra
potential $V=0$, then DKD leapfrog for equations (\ref{eq:ffff}) with
$\gamma=1$ can be written
\begin{eqnarray}
\bfr_\halff & = & \bfr+{\epsilon\mu\,\bfv\over v^2+2p_0}, \nonumber \\
t_\halff & = & t+ {\epsilon\mu\over v^2+2p_0}, \nonumber \\
\bfv' & = & \bfv- {\epsilon\mu\,\bfr_\halff\over r^2_\halff}, \nonumber \\
\bfr' & = & \bfr_\halff+{\epsilon\mu\,\bfv'\over (v')^2+2p_0}, \nonumber \\
t' & = & t_\halff + {\epsilon\mu\over (v')^2+2p_0},
\label{eq:dkdo} 
\end{eqnarray}
where $r=|\bfr|$, and $p_0$ is an integral of motion equal to minus the
initial energy. The fictitious time $\tau$ is equal to the eccentric anomaly
$u$ to within a linear transformation; each step of the integration
corresponds to $\Delta u=\epsilon(\mu/a)^{1/2}$ where $a=-\half\mu/E$ is the 
semimajor axis.

More generally, if the attracting mass has a trajectory $\bfr_\star(t)$ DKD
leapfrog with $\gamma=1$ reads
\begin{eqnarray}
\bfr_\halff & = & \bfr+{\epsilon\mu\bfv\over v^2+2p_0}, \nonumber \\
t_\halff & = & t+ {\epsilon\mu\over v^2+2p_0}, \nonumber \\
\bfv' & = & \bfv- {\epsilon\mu[\bfr_\halff-\bfr_\star(t_\halff)]
       \over |\bfr_\halff-\bfr_\star(t_\halff)|^2}, \nonumber \\
p_0' & = & p_0+ {\epsilon\mu[\bfr_\halff-\bfr_\star(t_\halff)]\cdot
d\bfr_\star(t_\halff)/dt
       \over |\bfr_\halff-\bfr_\star(t_\halff)|^2}, \nonumber \\
\bfr' & = & \bfr_\halff+{\epsilon\mu\bfv'\over (v')^2+2p_0'}, \nonumber \\
t' & = & t_\halff + {\epsilon\mu\over (v')^2+2p_0'}.
\label{eq:dkd} 
\end{eqnarray}

An appealing feature of equations (\ref{eq:dkd}) is that they contain no
square roots, which are the most time-consuming operation in
integrating Keplerian orbits by conventional methods; however, square root
evaluations do become necessary with this integrator as soon as the
non-Keplerian potential $V(\bfr,t)$ is non-zero.

\subsection{Error analysis for the Kepler problem}

\label{sec:errkep}

To analyze the numerical error of DKD leapfrog with $\gamma={3\over2}$ we take 
$f(x)$ from equation (\ref{eq:fdef}) and set $\Delta\tau=1$ in equation 
(\ref{eq:gamerr}). The leading term of the error Hamiltonian becomes
\begin{eqnarray}
\lefteqn{\Gamma_{\rm err}(\bfQ,\bfP) = {\epsilon^3\mu^3\over
24[-U(\bfq)T_e(\bfP)]^3}\times} \nonumber \\
& &\left\{2T_e^{3/2}|\bnabla U|^2
-p_ip_j(-U)^{3/2}U_{,ij}-[\ffrac{3}{2}(-U)^{1/2}+3T_e^{1/2}]
(\bfp\cdot\bnabla U)^2\right\}.
\end{eqnarray}
For the Kepler problem, $U=-\mu/r$ and the error Hamiltonian simplifies to
\be
\Gamma_{\rm err}={\epsilon^3\mu^2\over 24 T_e^3}
\left[{2T_e^{3/2}\over r}-{3T_e^{1/2}(\bfv\cdot\bfr)^2\over r^3}+
{3\mu^{1/2}(\bfv\cdot\bfr)^2\over 2r^{7/2}} - {\mu^{1/2}v^2\over
r^{3/2}}\right], 
\label{eq:errtt}
\ee
where $T_e=\half v^2+p_0$. 

Similarly, for $\gamma=1$ the leading term of the error Hamiltonian is
\be
\Gamma_{\rm err}(\bfQ,\bfP)={\epsilon^3\mu^3\over
24[-U(\bfq)T_e(\bfP)]^2}\left[2T_e|\bnabla U|^2+p_ip_jUU_{,ij}
-3(\bfp\cdot\bnabla U)^2\right];
\label{eq:errff}
\ee
for the Kepler problem, this simplifies to 
\be
\Gamma_{\rm err}(\bfQ,\bfP)={\epsilon^3\mu^3p_0\over
12r^2(\half v^2+p_0)^2}.
\label{eq:errkepham}
\ee
This formula leads to a remarkable conclusion, specific to this potential and
integrator. The original phase-space variables $(\bfq,\bfp)$ enter 
$\Gamma_{\rm err}$ in the same combination that they enter the Hamiltonian 
$\Gamma$; in other words the surrogate Hamiltonian may be written
\be
\widetilde\Gamma(\bfQ,\bfP)=\Gamma(\bfQ,\bfP)+\Gamma_{\rm err}(\bfQ,\bfP)
=\epsilon\mu\log W(\bfQ,\bfP)+{\epsilon^3\mu p_0\over 12
W^2(\bfQ,\bfP)},
\ee
where $W(\bfQ,\bfP)=r(\half p^2+p_0)/\mu$. Thus the equations of motion 
(\ref{eq:hamilton}) for $\widetilde\Gamma$ read
\be
\dot\bfz=\{\bfz,\widetilde\Gamma\}=\left({\epsilon\mu\over W}-{\epsilon^3\mu
p_0\over 6W^3}\right)\{\bfz,W\}, \quad \dot q_0={\epsilon r\over
W}-{\epsilon^3rp_0\over 6W^3}
+{\epsilon^3\mu\over 12 W^2}, \quad p_0=\hbox{const};
\ee
while the equations of motion for $\Gamma$ read
\be
\dot\bfz=\{\bfz,\Gamma\}={\epsilon\mu\over W}\{\bfz,W\}, 
\qquad \dot q_0={\epsilon r\over
W}, \qquad p_0=\hbox{const}.
\ee
Thus the exact trajectory $\bfz(\tau)$ (Hamiltonian $\Gamma$) is the same as
the numerical trajectory $\bfz(\tau')$ (Hamiltonian $\widetilde\Gamma$) at a
slightly different fictitious time, where the two timescales are related by
$d\tau=d\tau'[1-\epsilon^2p_0/(6W^2)]$ or $\tau=\tau'(1-\epsilon^2p_0)$ since
$W=1$ on the trajectory. In other words the algorithm (\ref{eq:dkd}) follows
the Keplerian trajectory {\it exactly}: the position and velocity are
precisely those of the Keplerian orbit, and the only error is in the time of
arrival at a given location (i.e. the only error is
``along-track")\footnote{This result holds only for DKD leapfrog, not KDK
leapfrog.}. Although we have only established this result to leading order in
the error Hamiltonian, we show in the Appendix that it holds at all orders,
i.e. for arbitrarily large timesteps. This result also applies in the more
general case where the attracting mass is in uniform motion rather than
stationary at the origin (cf. eqs. \ref{eq:dkd}).

We also show in the Appendix that  the timing error arising from a step
$\Delta u$ in eccentric anomaly is ${1\over 12}(\Delta u)^3/n+\hbox{O}(\Delta
u)^5$ where $n=(\mu/a^3)^{1/2}$ is the mean motion and the error is
independent of eccentricity. The fractional error in the orbital period is
then $\pi^2/3N^2$ where $N$ is the number of steps per period
(eq. \ref{eq:error}).

\subsection{Error analysis for the perturbed Kepler problem}

\label{sec:pertkep}

Because DKD leapfrog with $\gamma=1$ follows a Keplerian trajectory exactly,
this method is of particular interest for the perturbed Kepler problem, where
the extra potential $V(\bfr,t)$ in equations (\ref{eq:ffff}) is small but
non-zero. To investigate the errors in this case, we take 
the error Hamiltonian (\ref{eq:errff}), set $U=-\mu/r+V$ (for simplicity we 
assume that $V$ is stationary), and expand to first order in $V$:
\begin{eqnarray}
\lefteqn{\Gamma_{\rm err}(\bfQ,\bfP)= {\epsilon^3\mu^2\over 
24(\half v^2+p_0)^2}\bigg[{2\mu p_0\over r^2} + {4p_0V\over r} 
+4(\half v^2+p_0){\bfr\cdot\bnabla V\over r}} \nonumber \\
\qquad & &\mbox{}-rv_iv_jV_{,ij}+{v^2V\over r}-{3(\bfv\cdot\bfr)^2V\over r^3}
-6{\bfv\cdot\bfr\over r}\bfv\cdot\bnabla V\bigg].
\label{eq:errffff}
\end{eqnarray}
When this is evaluated on the trajectory, we have
\begin{eqnarray}
\lefteqn{\Gamma_{\rm err}(\bfQ,\bfP)= -\ffrac{1}{12}\epsilon^3\mu E+
{\epsilon^3\over 24}\bigg[-8ErV +4\mu\bfr\cdot\bnabla V}
\nonumber \\
 & & \mbox{}-r^3v_iv_jV_{,ij}+rv^2V-{3(\bfv\cdot\bfr)^2V\over r}
-6r(\bfv\cdot\bfr)\bfv\cdot\bnabla V\bigg].
\label{eq:errffffss}
\end{eqnarray}
The energy error is then (eq. \ref{eq:enerr})
\be
\Delta E={\Gamma_i-\Gamma_{\rm err}(\bfQ,\bfP)
\over \epsilon r}.
\label{eq:enerrone}
\ee
As $r\to 0$ we have $v\sim r^{-1/2}$. Thus if $V\sim r^k$ as $r\to 0$,
$\Gamma_{\rm err}\sim r^k$ as well. Then if $k>0$, the energy error at 
close encounters with the attracting mass is
\be
\Delta E={\Gamma_i\over \epsilon r}, \qquad r\ll r_i;
\label{eq:errstark}
\ee
in other words the energy error at close encounters is determined by the
initial conditions and varies as $r^{-1}$, independent of the form of the
perturbing potential at small radii so long as $V\to 0$ as $r\to 0$. 

The divergence of $\Delta E$ as $r\to 0$, even when the perturbing potential
$V\to0$ as $r\to 0$, appears to contradict our proof that the integrator
tracks Keplerian orbits exactly; the resolution is that the integrator only
tracks Keplerian orbits on the hypersurface $\Gamma=0$ in the extended phase
space, and numerical errors at larger radii perturb the trajectory to the
neighboring hypersurface $\Gamma=\Gamma_i$. In fact it can be shown that in
this case the integrator {\it is} following a Kepler orbit exactly, but for an
attracting mass $\mu\exp(\Gamma_i/\epsilon\mu)$. Thus even large energy errors
at close encounters do not signal a catastrophic failure of the integrator, so
long as $|\Gamma_i/\epsilon\mu|\ll 1$.

Moreover there is a simple way to correct these errors. Normally the initial
value of $p_0$ is set equal to $-E$, so that $\Gamma=0$; instead, we modify
the initial value of $p_0$ so that $\widetilde\Gamma=\Gamma+\Gamma_{\rm err}$
is zero. This requires
\be
p_0=-E+{\mu\over r}\left[\exp(-\Gamma_i/\epsilon\mu)-1\right].
\label{eq:correct}
\ee

\section{Numerical tests}

\label{sec:num}

\subsection{Keplerian two-body problem}

We have tested these integration methods by following Keplerian orbits with
eccentricities $e = 0.9$, $0.99$, $0.999$, $0.9999$, $0.99999$, and
$0.999999$. Each orbit is started at pericenter and followed for $2\times
10^4$ orbital periods (although a shorter integration would have been
sufficient, since the energy errors are oscillatory
rather than growing). We characterize the performance of the integrator by the
maximum energy error $|\Delta E_{\rm max}/E_0| \equiv \max |(E-E_{0})/E_{0}|$,
as a function of the number of steps per orbital period.

Figure \ref{fig:one} shows the energy error that arises from integrating
equations (\ref{eq:ffff}) using DKD leapfrog with $\gamma={3\over2}$
(i.e. timestep $\propto r^{3/2}$). For comparison
we have also shown as open circles the energy error for leapfrog with fixed
timestep ($\gamma=0$) at eccentricity $e=0.9$ and 0.99. Clearly
$\gamma={3\over2}$ provides far more accurate integrations than $\gamma=0$.

\begin{figure} \vspace{10cm} \includegraphics{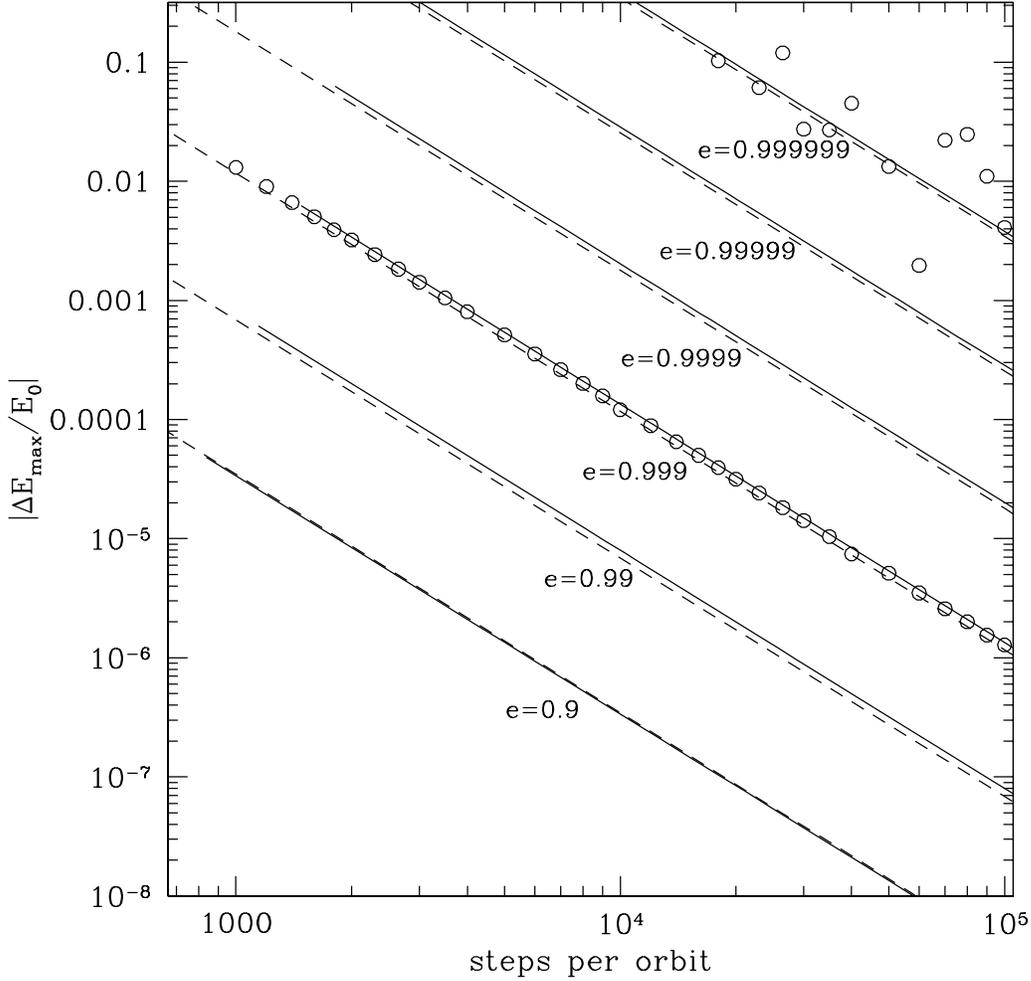} \caption{Maximum fractional energy error over
$2\times10^4$ orbital periods, as a function of number of steps
per orbit for the Keplerian two-body problem. The curves correspond to
eccentricities $e=0.9, 0.99, 0.999, 0.9999, 0.99999, 0.999999$. The integrator
is standard DKD (drift-kick-drift) leapfrog with timestep $\propto r^{3/2}$,
following equations (\ref{eq:gpow}) and (\ref{eq:ffff}) with
$\gamma={3\over2}$. The orbits are started at pericenter. 
The dashed lines show analytic estimates of the energy
error from equations (\ref{eq:perierr}) and (\ref{eq:nstep}). 
The analogous errors for DKD leapfrog with fixed timestep
and $e=0.9$ and $0.99$ are shown as open circles; for larger eccentricities the
fixed-timestep errors are off-scale.} \label{fig:one} \end{figure}

We can compare these energy errors to the analysis of \S
\ref{sec:errkep}. When evaluated on the trajectory ($T_e=-U$) the error
Hamiltonian (\ref{eq:errtt}) is 
\be 
\Gamma_{\rm err}=-{\epsilon^3r^{3/2}\over 24
\mu^{1/2}}\left[{3(\bfr\cdot\bfv)^2\over 2r^2}+2E\right]={\epsilon^3na^2\over
24}\left[(1-e\cos u)^{3/2}-{3e^2\sin^2u\over 2(1-e\cos u)^{1/2}}\right],
\ee
where as usual $n$, $a$, and $u$ are the mean motion, semimajor axis, and
eccentric anomaly. The energy error is then given by equation
(\ref{eq:enerr}),
\begin{eqnarray}
\Delta E & = &{\epsilon^2n^2a^2\over 24}\bigg[\left(1-e\cos u_i\over 1-e\cos
u\right)^{3/2} -{3e^2\sin^2 u_i\over2(1-e\cos u_i)^{1/2}(1-e\cos u)^{3/2}}
\nonumber \\
& {} &\qquad\qquad -1+{3e^2\sin^2 u\over 2(1-e\cos u)^2}\bigg].
\end{eqnarray}
For high-eccentricity orbits started at pericenter ($u_i=0$), the maximum
error $|\Delta E|$ occurs at $u\simeq \cos^{-1}e$, and is given by
\be
\left|\Delta E\over E\right|_{\rm max}
={\epsilon^2\over 16(1-e)} +\hbox{O}[\epsilon^2(1-e)^0].
\label{eq:perierr}
\ee
For high-eccentricity orbits started at apocenter ($u_i=\pi$), the maximum
energy error occurs at pericenter, 
\be
\left|\Delta E\over E\right|_{\rm max}=
{\epsilon^2\over3\cdot 2^{1/2}(1-e)^{3/2}}+
\hbox{O}\left(\epsilon^2\over 1-e\right).
\ee
These formulae show that pericenter starts lead to smaller energy errors than
apocenter starts, although in the latter case the errors can be reduced by the 
use of corrected initial conditions (cf. \S \ref{sec:pertkep}). 

The number of steps per orbit $N\simeq\int_0^Pdt/g(r)$, where $g(r)$
is the timestep function (\ref{eq:gpow}) and $P$ is the orbital period. For
$\gamma={3\over2}$
\be
N={2\over\epsilon}\int_0^\pi {df\over(1+e\cos f)^{1/2}}={4\over
\epsilon(1+e)^{1/2}}K\left(2e\over 1+e\right),
\label{eq:nstep}
\ee
where $f$ is the true anomaly and $K$ is an elliptic integral. Plotting
equations (\ref{eq:perierr}) and (\ref{eq:nstep}) as a parametric function of
$\epsilon$ we obtain the dashed lines in Figure \ref{fig:one}, which agree
well with the energy errors from the numerical orbit integrations.

As we discussed in the previous section, integrating the Keplerian equations of
motion using DKD leapfrog with $\gamma=1$ (eq. \ref{eq:dkdo}) yields even
better behavior than $\gamma={3\over2}$, as in this case the energy error is
zero---in fact there is zero error in all of the phase-space functions that
are constants of motion in a point-mass potential (energy, angular momentum
and Runge-Lenz vector).  To test the practical value of this algorithm we must
therefore turn to more general Hamiltonians, which we now do.

\subsection{The Stark problem}

The Stark problem is to follow the motion of a test particle subject to an
inverse-square force plus a constant force; the Stark Hamiltonian is 
\be
H=\half \bfp^2-{\mu\over r}-\bfS\cdot\bfr, 
\label{eq:stark}
\ee 
where the Stark vector $\bfS$ is a constant. The Stark Hamiltonian has three 
constants of motion and thus is integrable: these are the energy $E$,
the angular momentum component along $\bfS$, and a third analytic integral
that we see no point in writing out (\cite{Pa65,LL76}). We restrict ourselves
to the planar case, which is particularly challenging because all orbits
oscillate between retrograde and prograde and hence pass arbitrarily close to
the attracting mass. We shall examine only a single integrator, DKD leapfrog, 
applied to the equations of motion for $\gamma=1$ (eqs. \ref{eq:ffff}), 
since in this case the trajectory is followed exactly when $\bfS=0$.  

\begin{figure} \vspace{10cm} \includegraphics{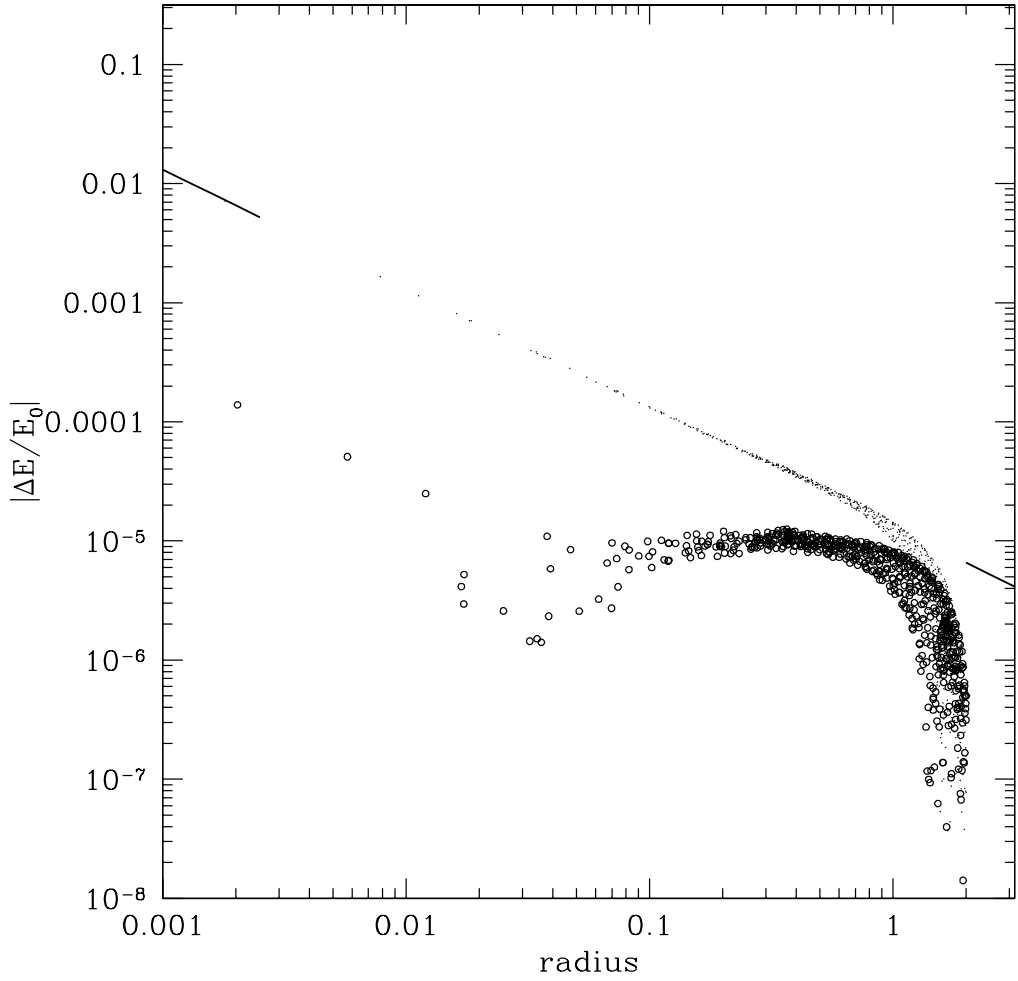} 
\caption{Fractional energy error as a function of
distance from the attracting mass, for a numerical integration of the Stark
problem using DKD leapfrog with $\gamma=1$, which integrates Keplerian orbits
with zero energy error. The integration lasts for 1000 Keplerian periods of the
initial orbit and the error is plotted every 100 timesteps. The integration
parameters are $\mu=1$, $\epsilon=0.1$, the initial
eccentricity is $e=0.9$, the Stark vector is $45^\circ$ from the initial line
of apsides, and its magnitude is $S=\eta E^2/\mu$ where $\eta=4\times10^{-3}$.
For $r\ll 1$ the points lie on a straight line, consistent with the
prediction of equation (\ref{eq:errstark}) marked by solid line segments.
The open circles show the much smaller 
energy errors when the initial conditions are corrected using equation 
(\ref{eq:correct}). }
\label{fig:two}
\end{figure}

Rauch \& Holman (1999) \nocite{RH99} have recently tested several integrators
on the Stark problem, and we shall usually use their initial conditions: the
initial eccentricity $e=0.9$, the Stark vector is oriented $45^\circ$ to
the initial line of apsides, and the orbit is started at apocenter. The
strength of the Stark perturbation is written $S=\eta E^2/\mu$ where $E$ is
the energy and $\eta\ll1$ for nearly Keplerian motion.

The error Hamiltonian and the expected energy error are given by equations
(\ref{eq:errffffss}) and (\ref{eq:enerrone}) with $V=-\bfS\cdot\bfr$.  Figure
\ref{fig:two}\ verifies the functional form $\Delta E\propto r^{-1}$ predicted
by (\ref{eq:enerrone}) and demonstrates the improvement during close
encounters that results from using the corrected initial condition
(\ref{eq:correct}).

Figure \ref{fig:three} shows the energy error as a function of steps per orbit
$N$ and strength of the Stark parameter, for integrations lasting $n=10^4$
orbital periods using the Rauch-Holman initial conditions corrected by
equation (\ref{eq:correct}).  We have plotted both the maximum energy error,
which is dominated by very close encounters, and the average of the absolute
value of the energy error, which provides a better estimate of the typical
error. The average error exhibits the $N^{-2}$ dependence expected for a
second-order integrator; the maximum error is much larger and more irregular,
reflecting its dependence on rare and rather unphysical close encounters (the
typical maximum eccentricity in an integration of this length is given by
$1-e_{\rm max}\sim n^{-2}=10^{-8}$). We have also plotted one error curve for
uncorrected initial conditions; we see that the correction reduces the errors
by about one order of magnitude in this case.

\begin{figure} \vspace{10cm} \includegraphics{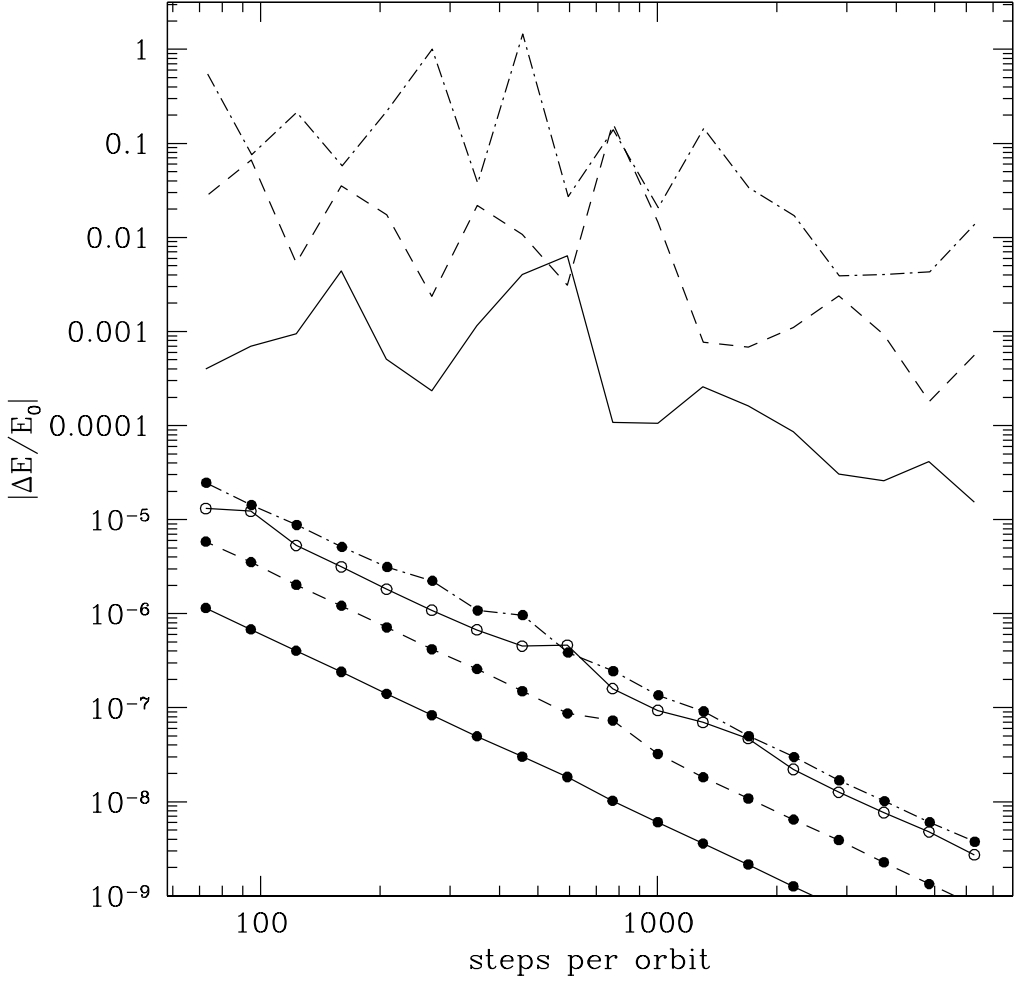} 
\caption{Fractional energy error for the Stark problem, as a function of
number of force evaluations per orbit. The initial eccentricity is $e=0.9$,
the Stark vector is $45^\circ$ from the initial line of apsides, and the orbit
starts at apocenter and is followed for $10^4$ periods. The magnitude of the
Stark vector is $S=\eta E^2/\mu$ where $\eta=0.001$ (solid lines), 0.005
(dashed lines), and 0.02 (dash-dot lines). The lower curves (solid circles) 
represent the average of the absolute value of the energy error, and the 
upper curves the maximum error. The integrator is DKD leapfrog with 
$\gamma=1$ and initial conditions corrected using equation (\ref{eq:correct}); 
the single solid curve with open circles represents the average error for
$\eta=0.001$ if no corrector is applied.} \label{fig:three}
\end{figure}

Rauch \& Holman (1999) tested the Wisdom \& Holman (1991) \nocite{WH91}
integrator on the Stark problem with $\eta=4\times10^{-3}$. They found that
the Wisdom-Holman integrator---which works very well for low-eccentricity
orbits---was generally unstable, in that the energy error grew by a random
walk until the orbit escaped to infinity. The instability arose through
numerical chaos caused by the overlap of resonances in the error Hamiltonian,
and could only be evaded if the (constant) timestep was in resonance with the
orbital period, or is small enough that the pericenter passage is
well-resolved---even though the Wisdom-Holman integrator follows Keplerian
orbits exactly for any timestep. Our integrator is evidently not subject to
these limitations.

Rauch \& Holman also tested several other methods. In particular, Mikkola's
regularized version of the Wisdom-Holman mapping was completely stable, and
gave energy errors comparable to those shown for our integrator in Figure
\ref{fig:three}. However, we expect that our method is faster in practice
because it requires fewer calculations per integration step. 

\section{Summary}

\label{sec:disc}

We have constructed adaptive-timestep, reversible, explicit \sias\
for separable Hamiltonians of the form (\ref{eq:hamsep}), using extended phase
space (\cite{Mi97}); the principal restriction is that the timestep must be a
function of the potential energy alone.

Integrators of this kind would require modifications for problems with many
degrees of freedom since the total potential energy of the system is
insensitive to local conditions that may demand a short timestep (e.g. a close
encounter between two bodies). However, for test-particle integrations in
fixed, smooth potentials or few-body systems with similar masses, these
integrators can provide both adaptive timestep control and the excellent
long-term error control associated with symplectic and reversible integration
algorithms.

For close encounters or eccentric orbits in few-body gravitating systems,
these adaptive-timestep \sias\ provide an attractive alternative to
fixed-timestep \sias\ in regularized coordinates; moreover, unlike regularized
integrators, adaptive-timestep \sias\ can also be used to follow 
orbits in non-Keplerian potentials (e.g. galaxies).

Although we have discussed only second-order leapfrog integrators,
higher-order adaptive-timestep \sias\ can be derived by concatenating leapfrog
steps of different lengths (\cite{Yo90}).

A particularly interesting example of these integrators is offered by
equations (\ref{eq:dkdo}), which follow a Keplerian trajectory exactly, with
only ``along-track" errors.

\bigskip

This research was supported in part by NASA Grant NAG5--7310. We thank Seppo
Mikkola for sending us a copy of his paper (\cite{MT99}), which also shows
that the integrator (\ref{eq:dkd}) is exact for Keplerian orbits. We also
thank Kevin Rauch for thoughtful comments. 

\newpage\appendix
\section{Appendix: An exact integrator for the Kepler problem} 

We prove that the $\gamma=1$ leapfrog integrator (\ref{eq:dkd}) follows a
Keplerian orbit exactly, except for errors in the time. For simplicity we
shall assume that the attracting mass is at rest at the origin, as in
equations (\ref{eq:dkdo}); the extension to an attracting mass in uniform
motion is straightforward.

Let $(\bfr,\bfv)$ and $(\bfr',\bfv')$ be the position and velocity at two
points on a bound Keplerian orbit with eceentric anomalies $u$ and $u'$ 
respectively. Then $(\bfr',\bfv')$ satisfies the relation
\begin{eqnarray}
\bfr' & = & f(u,u')\bfr + g(u,u') \bfv  \nonumber \\
\bfv' & = & f_{t}(u,u')\bfr + g_{t}(u,u') \bfv,  
\end{eqnarray}
where 
\begin{eqnarray}
f(u,u') & = & {\cos(u'-u)-e\cos u\over 1-e\cos u}, \nonumber  \\
g(u,u') & = & {1\over n}\left[\sin(u'-u)-e\sin u' + e\sin u\right], 
\nonumber \\
f_{t}(u,u') & = & -{n\sin(u'-u)\over (1-e\cos u')(1-e\cos u)},  \nonumber \\
g_{t}(u,u') & = & {\cos(u'-u)-e\cos u'\over 1-e\cos u'}
\end{eqnarray}
are Gauss's $f$ and $g$ functions (e.g. \cite{Da88}); here $e$ is the
eccentricity, $n=(\mu/a^3)^{1/2}$ is the mean motion and $a$ is the semimajor 
axis, which is related to the radius through $r=a(1-e\cos u)$. 

These equations can be rewritten as
\begin{eqnarray}
\bfr_\halff & = & \bfr + s(u',u)\bfv, \nonumber \\
\bfv' & = & \bfv +z(u',u)\bfr_\halff, \nonumber \\
\bfr' & = & \bfr_\halff - s(u,u')\bfv',
\label{eq:mmmm}
\end{eqnarray}
where
\begin{eqnarray}
s(u',u) & = & {[1-\cos(u'-u)](1-e\cos u)\over n\sin(u'-u)}, \nonumber \\
z(u',u) & = & -{n\sin(u'-u)\over (1-e\cos u')(1-e\cos u)}.
\end{eqnarray}.

Comparison to the $\gamma=1$ leapfrog integrator (\ref{eq:dkd}) shows that the
two maps $(\bfr,\bfv)\to(\bfr',\bfv')$ are the same if
\begin{eqnarray}
{\epsilon\mu\over p^2+2p_0} & = & {[1-\cos(u'-u)](1-e\cos u)\over n\sin(u'-u)},
\nonumber \\
{\epsilon\mu\over q_\halff^2} & = & {n\sin(u'-u)\over(1-e'\cos u')(1-e\cos
u)}, \nonumber \\
{\epsilon\mu\over (p')^2+2p_0} & = & {[1-\cos(u'-u)](1-e\cos u')\over 
n\sin(u'-u)}.
\label{eq:qqqq}
\end{eqnarray}
We use the relations $p^2+2p_0=2\mu/[a(1-e\cos u)]$, 
$(p')^2+2p_0=2\mu/[a(1-e\cos u')]$ and square the first of equations
(\ref{eq:mmmm}) to eliminate $q_\halff^2$. After some algebra we find that 
all of the relations (\ref{eq:qqqq}) are satisfied if
\be
\epsilon=2{1-\cos\Delta u\over na\sin\Delta u}.
\label{eq:cond}
\ee
where $\Delta u\equiv u'-u$. Thus, we have proved that our mapping
follows the Keplerian two-body problem exactly in the original phase
space. Although our proof is for bound orbits, it is straightforward to show
that unbound Keplerian orbits ($a<0$) are also integrated exactly, with
\be
\epsilon=2{\cosh\Delta u-1\over n_ua\sinh\Delta u};
\label{eq:condu}
\ee
here $n_u=(-\mu/a^3)^{1/2}$ and $r=-a(e\cosh u-1)$.

We must still establish the relation between the timestep given by
the second and fifth of equations (\ref{eq:dkd}) and the actual time $\Delta
t_K$ required to travel from $\bfr$ to $\bfr'$. The timestep is
\be 
t'=t+\Delta t=t+\epsilon\mu\left({1\over
p^2+2p_0}+{1\over (p')^2+2p_0}\right)=t+\half\epsilon a(2+e\cos u+e\cos u').  
\ee 
The relation (\ref{eq:cond}) then implies that 
\begin{eqnarray} 
n\Delta t & = & {[1-\cos(u'-u)](2-e\cos u-e\cos u')\over \sin(u'-u)}\nonumber
\\ & = & 2{1-\cos(u'-u)\over \sin(u'-u)}-e\sin u'+e\sin u.  
\end{eqnarray}
On the other hand Kepler's equation states that the actual timestep is given by
\be
n\Delta t_K=u'-u-e\sin u'+e\sin u.
\ee
Thus the time error is given by
\be
n(\Delta t-\Delta t_K)= 2{1-\cos\Delta u\over \sin\Delta u}-\Delta u=
\ffrac{1}{12}(\Delta u)^3+ \ffrac{1}{120}(\Delta u)^5+\hbox{O}(\Delta u)^7,
\ee
independent of eccentricity. 

If we take $N=2\pi/\Delta u$ steps per orbit, the timing error per orbit is
\be
{\delta t\over P}={\pi^2\over 3N^2}+\hbox{O}(N^{-4}),
\label{eq:error}
\ee
independent of eccentricity.

\end{document}